\documentclass[prx,twocolumn,notitlepage,superscriptaddress,showpacs,longbibliography]{revtex4-1}

\usepackage{amsmath}
\usepackage{amsfonts}
\usepackage{amsthm}
\usepackage[utf8]{inputenc}
\usepackage{graphicx}
\usepackage{color}
\usepackage{quantum}
\usepackage[colorlinks]{hyperref}
	\hypersetup{
		bookmarksnumbered,
		pdfstartview={FitH},
		citecolor={blue},
		linkcolor={red},
		urlcolor={black},
		pdfpagemode={UseOutlines}
	}

\newcommand{\mc}[1]{\mathcal{#1}}

\newcommand{\dmin}{\delta_\text{min}}

\newtheorem{lem}{Lemma}

\newtheorem{res}{Result}
\newtheorem{defn}{Definition}

\begin{document}
\title{Quantum processes which do not use coherence}

\author{Benjamin Yadin} \email{benjamin.yadin@physics.ox.ac.uk}
\affiliation{Atomic and Laser Physics, Clarendon Laboratory, University of Oxford, Parks Road, Oxford, OX1 3PU, UK}

\author{Jiajun Ma}
\affiliation{Center for Quantum Information, Institute for Interdisciplinary Information Sciences, Tsinghua University, 100084 Beijing, China}

\author{Davide Girolami} \email{davegirolami@gmail.com}
\affiliation{Atomic and Laser Physics, Clarendon Laboratory, University of Oxford, Parks Road, Oxford, OX1 3PU, UK}

\author{Mile Gu} \email{ceptryn@gmail.com}
\affiliation{School of Physical and Mathematical Sciences, Nanyang Technological University, Singapore 639673, Republic of Singapore}
\affiliation{Complexity Institute, Nanyang Technological University, 60 Nanyang View, Singapore 639673, Republic of Singapore}
\affiliation{Centre for Quantum Technologies, National University of Singapore, Singapore 117543}

\author{Vlatko Vedral}
\affiliation{Atomic and Laser Physics, Clarendon Laboratory, University of Oxford, Parks Road, Oxford, OX1 3PU, UK}
\affiliation{Center for Quantum Information, Institute for Interdisciplinary Information Sciences, Tsinghua University, 100084 Beijing, China}
\affiliation{Centre for Quantum Technologies, National University of Singapore, Singapore 117543}
\affiliation{Department of Physics, National University of Singapore, 2 Science Drive 3, 117551 Singapore}

\date{\today}
\begin{abstract}
A major signature of quantum mechanics beyond classical physics is coherence, the existence of superposition states. The recently developed resource theory of quantum coherence allows the formalisation of incoherent operations -- those operations which cannot create coherence. We identify the set of operations which additionally do not use coherence. These are such that coherence cannot be exploited by a classical observer, who measures incoherent properties of the system, to go beyond classical dynamics. We give a physical interpretation in terms of interferometry and prove a dilation theorem, showing how these operations can always be constructed by interacting the system in an incoherent way with an ancilla. Such a physical justification is not known for the incoherent operations, thus our results lead to a physically well-motivated resource theory of coherence. Next, we investigate the implications for coherence in multipartite systems. We show that quantum correlations can be defined naturally with respect to a fixed basis, providing a link between coherence and quantum discord. We demonstrate the interplay between these two quantities under our studied operations, and suggest implications for the theory of quantum discord by relating the studied operations to those which cannot create discord.
\end{abstract}

\maketitle

\section{Introduction}
Quantum technologies promise to deliver an advantage over their classical counterparts in a diverse set of tasks ranging from computation to high-precision metrology to heat engines. In recent years, much effort has been directed towards identifying the quantum resources necessary for an increased performance in these tasks. However, it is not always easy to unambiguously define the classical analogue of a given quantum protocol (if it exists); similarly, many nonclassical signatures of quantum mechanics have been developed, including entanglement, discord and contextuality.

Here, we focus on one of the most fundamental quantum features: coherence, or superposition. We suppose that a given system naturally admits a classical description in a certain preferred basis. For instance, a charge transport network has classical states in which the charged particle is localised at one of the sites. Nonclassicality is then associated with superpositions of these states.

This is the view taken by the recently developed resource theory of coherence \cite{baumgratz2014quantifying}. The resource theory approach has proved to be highly useful in many areas of quantum information theory, including entanglement, thermodynamics and asymmetry \cite{horodecki2013quantumness,brandao2013resource,gour2008resource,brandao2015reversible}. For the coherence resource theory, the states which are diagonal in a preferred basis are chosen as the free states which can be prepared with no resource cost. These are known as incoherent states. In order to identify measures of coherence, one also needs to define a set of free operations, from which one stipulates that a good measure cannot increase under free operations. This gives axiomatic criteria for coherence measures. The original work used the set of incoherent operations, defined such that they can never create coherence from an incoherent state.

There have been a number of recent works investigating coherence as resource and its manipulation under incoherent operations \cite{winter2016operational,du2015conditions,chitambar2015assisted,chitambar2015relating,streltsov2015measuring,streltsov2015hierarchies,hu2015coherence,girolami2015witnessing,lostaglio2015description,lostaglio2015quantum}. By now, a variety of different candidates for the free operations have been proposed \cite{marvian2016quantify,devicente2016power,chitambar2016critical} due to the lack of a general experimental setting where coherence is a resource. Here, we propose a characterisation of free incoherent operations from physical considerations.

We suggest a set of incoherent operations with a new restriction, which we term as the inability to use coherence. We assume that the relevant outputs from a process are classical properties -- those accessible to a classical observer who is limited to measurements in the incoherent basis. Then these operations are such that the outputs are independent of the coherence of the system.

Crucially, the outputs of such a process can be fully described by a classical stochastic operation on the input probability distribution. Thus these processes can never outperform classical ones, as far as a classical observer is concerned. Coherence may be present, but it does not contribute to the task.

We first give a formal definition of the processes that can neither create nor use coherence, and characterise them as the \emph{strictly incoherent} (SI) operations defined in Ref.\ \cite{winter2016operational}. Motivated by interferometry, we give a prescription for how these operations are generated by incoherent interactions with an ancilla.  Our results suggest that the abilities to prepare and to detect coherent states can both be seen as resources. No such physical picture has yet been presented for the incoherent operations. Hence the SI operations are a physically well motivated set of free operations for coherence. We also give a set of coherence measures which are monotones under SI operations but not necessarily under all incoherent operations.

Next, we study the implications of our results for multipartite systems. In such a setting, it has long been held that correlations, and in addition quantum correlations, can be useful resources. While the interplay between coherence and entanglement has been studied recently \cite{chitambar2015assisted,chitambar2015relating,streltsov2015hierarchies}, we are concerned here with a more general kind of quantum correlation called quantum discord \cite{ollivier2001quantum,henderson2001classical,modi2012classical}. Some works have begun to show fundamental links between coherence and discord \cite{bromley2015frozen,yao2015quantum,ma2015converting}.

While some operational interpretations of discord are known \cite{gu2012observing,madhok2011interpreting,cavalcanti2011operational,dakic2012quantum}, its status as a resource lacks a firm footing since there is no known associated resource theory. Part of improving this situation will involve understanding the behaviour of discord under local operations. One significant consequence is that discord can increase under local operations \cite{streltsov2011behavior} -- which is counterintuitive for a measure of quantum correlations.

Here, we identify a refined form of discord which we show measures the quantumness of correlations with respect to the incoherent basis. This is often overlooked in favour of the usual basis-independent form -- we highlight basis-dependent discord as a quantity of significance by demonstrating its strong connections with coherence. We characterise its behaviour under local SI operations, finding conditions under which it can be created or consumed. We show that the particular structure of the SI operations enables an interpretation of its behaviour via the manipulation of classical information. In particular, an increase in this discord can be attributed to a loss of a classical memory recording which operation was performed. Furthermore, we suggest a new set of basis-dependent discord measures with exactly the same behaviour with respect to SI operations, quantifying the loss of nonlocal information under local dephasing.

Finally, we find a new fundamental connection between quantum coherence and basis-independent discord. Namely, those operations which are SI in every basis, combined with unitary operations, form a significant subset of all those operations which cannot create discord. Identifying the free operations for discord is an open problem, and this provides the first hint towards a solution.

\section{Defining the operations}

We shall study operations which have two properties: they can neither create nor use coherence. To define the former property, we review the resource theory of coherence. For a state space of dimension $d$, one chooses a preferred incoherent basis $\{\ket{i}\},\, i=0,1,\dots,d-1$; an incoherent state is then any mixture of these, $\sum_i p_i \ketbra{i}{i}$. An operation which cannot create coherence (i.e., an incoherent operation) is such that, when an incoherent state is input, the output must be another incoherent state. 

\begin{figure}[h]
	\centering
	\includegraphics[scale=.85]{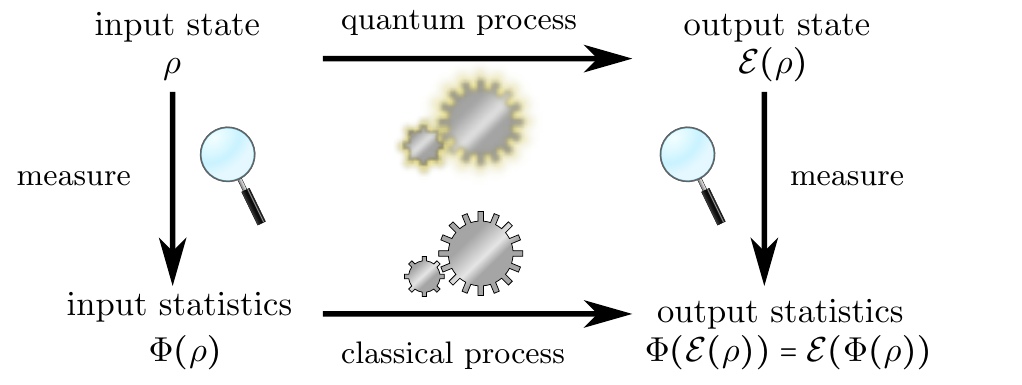}
	\caption{Incoherent operations cannot create coherence; strictly incoherent operations in addition cannot use coherence. We depict here an SI operation $\mc{E}$, showing that all the initial coherence in the input $\rho$ can be removed with a measurement $\Phi$ (in the incoherent basis) without affecting the final measurement outcomes.}
	\label{fig:process}
\end{figure}

For the latter property, we consider a case where the outputs of a protocol $\rho \to \sigma$ are determined by measurements performed by a classical observer in the incoherent basis -- in other words, the probabilities $\braXket{i}{\sigma}{i}$. These probabilities must then be the same regardless of whether the input state is $\rho$ or the ``dephased" state $\Phi(\rho) := \sum_i \braXket{i}{\rho}{i} \ketbra{i}{i}$, which has no coherence but the same diagonal elements as $\rho$. This is depicted in Fig. \ref{fig:process}.

The assumption of incoherent outputs is relevant in situations such as transport mechanisms, where one is interested in the probability that an excitation is located at a specific site after a certain amount of time \cite{huelga2013vibrations,levi2014quantitative}. Identifying the localised state at site $i$ with $\ket{i}$, this probability is then $p_i=\braXket{i}{\rho}{i}$. For an operation which does not use coherence, $p_i$ for the output state remains unchanged when the input coherence is removed by $\Phi$.

To formalise this, let us first recall that a general quantum operation $\mc{E}$ can be defined in the framework of completely positive maps by a set of Kraus operators $\{K_\mu\}$ such that its action on any state $\rho$ is $\mc{E}(\rho) = \sum_\mu K_\mu \rho K_\mu^\dagger$. Each $K_\mu$ is associated with a selected measurement outcome $\rho^\mu = K_\mu \rho K_\mu^\dagger / p_\mu$ with probability $p_\mu = \tr(K_\mu \rho K_\mu^\dagger)$, and the requirement that $\sum_\mu K_\mu^\dagger K_\mu \leq I$ ensures that the probabilities are (sub)normalised. There is generally no unique choice of Kraus operators.

Following Ref.\ \cite{baumgratz2014quantifying}, we define an incoherent Kraus operator to always map incoherent states to incoherent states: that is, 
\begin{equation} \label{eqn:incoherent_kraus}
	K_\mu \ket{i} \propto \ket{f_\mu(i)}
\end{equation}
for some function $f_\mu$. An incoherent operation $\mc{E}$ then has some set of Kraus operators which are all incoherent \footnote{We note, as in Ref. \cite{winter2016operational}, that this is stronger than just imposing that $\mc{E}$ takes the set of incoherent states to itself, as every selected outcome must be an incoherent operation.}. Our additional requirement on the operations is:

\begin{defn}
	An operation $\mc{E}$ is said to \emph{not use coherence} if and only if it has a set of incoherent Kraus operators $\{K_\mu\}$ such that measurement outcomes in the incoherent basis are independent of the coherence of the input state:
	\begin{equation} \label{eqn:si_definition}
		\braXket{i}{K_\mu \rho K_\mu^\dagger}{i} = \braXket{i}{K_\mu \Phi(\rho) K_\mu^\dagger}{i} \quad \forall \mu, i.
	\end{equation}
\end{defn}

A concise way of stating this condition on $K_\mu$ is that the operation $\mc{E}^\mu: \rho \to K_\mu \rho K_\mu^\dagger$ commutes with dephasing:
\begin{equation}
	[\mc{E}^\mu, \Phi] = 0,
\end{equation}
meaning that $\mc{E}^\mu(\Phi(\rho)) = \Phi(\mc{E}^\mu(\rho))$ for every state $\rho$ \footnote{This may be compared with the classical operations as defined in Ref.\ \cite{meznaric2013quantifying}}.

For such an operation, the transformation induced on the probabilities $p_i = \braXket{i}{\rho}{i}$ is $p_i \to \sum_j T_{i,j} p_j$, where $T$ is a (sub)stochastic matrix, ensuring the (sub)normalisation of the $p_i$. This represents a classical stochastic process.

These operations are obtained by choosing the $f_\mu$ in (\ref{eqn:incoherent_kraus}) to be invertible functions -- i.e., permutations of the set $\{0,1,\dots,d-1\}$. This coincides with the strictly incoherent (SI) operations defined in Ref.\ \cite{winter2016operational}. We say that the Kraus operator $K_\mu$ is SI when the operation $\rho \to K_\mu \rho K_\mu^\dagger$ is SI.

$K_\mu$ is SI exactly when both $K_\mu$ and $K_\mu^\dagger$ are incoherent. The simplest way to see this is to write $K_\mu^\dagger \ket{i} = \sum_j {c^\mu_{i,j}}^* \ket{j}$ and thus $\braXket{i}{K_\mu \rho K_\mu^\dagger}{i} = \sum_{j,j'} c^\mu_{i,j} {c^\mu_{i,j'}}^* \braXket{j}{\rho}{j'}$. The off-diagonal terms with $j \neq j'$ on the right-hand side always vanish as long as $c^\mu_{i,j}$ has at most one nonzero value of $j$ for each $i$. This is the condition for $K_\mu^\dagger$ to be incoherent.

\section{Physical implementation}
\subsection{Operations in an interferometer}
\begin{figure}[h]
	\centering
	\includegraphics[scale=.8]{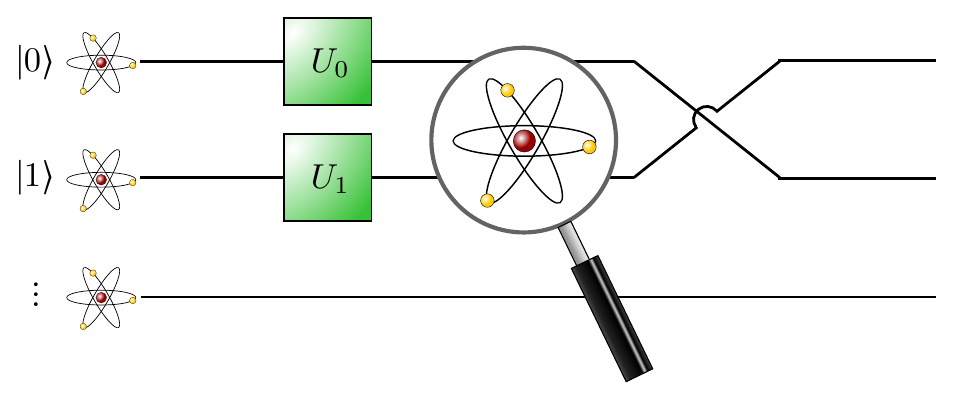}
	\caption{Strictly incoherent operations viewed as operations on a particle travelling through an interferometer. The different branches $\ket{i}$ span the Hilbert space $\mc{H}_B$, while the internal state of the particle has Hilbert space $\mc{H}_I$. The operations are built up by combining path-dependent unitary operations $U_i$ on and measurements of the internal state of the particle, and permutations of the paths.}
	\label{fig:interferometer}
\end{figure}

We now give a simple setting where the physical interpretation of the operations which neither create nor use coherence is particularly clear. Consider a particle passing through an interferometer with an arbitrary discrete number of branches. The Hilbert space of the particle can be written as $\mc{H} = \mc{H}_B \ox \mc{H}_I$, where $\mc{H}_B$ is the branch degree of freedom and $\mc{H}_I$ is the internal state of the particle. Interferometry relies on coherence between branches, so we associate each branch $i$ with an incoherent basis state ${\ket{i}}_B$. As we prove in the next section, the SI operations in this situation are those operations resulting from combining path-dependent unitaries on $\mc{H}_I$, measurements on $\mc{H}_I$, and permutations of the paths $i$. See Fig.\ \ref{fig:interferometer} for an illustration.

Recall that any interferometric protocol starts and ends with a beam splitter operation, with path-dependent phase gates in between. The function of the first beam splitter is to create coherence between branches - i.e., a superposition of the $\ket{i}$. The second recombines the branches such that subsequent which-path measurements reveal some information about the transformation induced by the phase gates. In other words, the second beam splitter enables measurement in a coherent basis.

This observation gives an intuition for the difference between SI and the full set of incoherent operations. When restricted to unitary operations, they are the same. For a simple example of an operation that is incoherent but not SI, consider the Kraus operators
\begin{equation} \label{eqn:non_si_example}
	K_0 = \ketbra{0}{+}, K_1 = \ketbra{1}{-}
\end{equation}
acting on a pair of branches $\ket{0}$ and $\ket{1}$, with $\ket{\pm} = (\ket{0} \pm \ket{1})/\sqrt{2}$. This is a measurement in the coherent basis $\{ \ket{\pm}\}$, such that the inputs $\ket{+}$ and $\ket{-}$ are mapped onto $\ket{0}$ and $\ket{1}$ respectively. By contrast, these two states are completely indistinguishable by SI operations since they have identical statistics in the incoherent basis.

More generally, within the SI framework, the ability to perform coherent measurements -- i.e., to measure in a basis $\{\ket{\psi_i}\}$ different from the incoherent basis -- constitutes an additional resource. In the resource theory of coherence defined by incoherent operations, these measurements are given for free (provided that the outputs are mapped to incoherent states). Although the resulting operation is incoherent, such a measurement in an interferometer requires use of a beam splitter -- a device capable of creating coherence. Thus it is natural in this context to regard incoherent but non-SI operations as being as difficult to perform as general coherent operations. We formalise this statement in the following section.

\subsection{Unitary interaction with an ancilla}

The inability of SI operations to make use of coherence can be also appreciated in their unitary dilation.  A fundamental theorem about quantum operations says that they can always be constructed by interacting the system $S$ in a unitary way with an ancilla $\alpha$ \cite{nielsen2010quantum}. For instance, any operation corresponding to a single Kraus operator $K$ can be constructed by interacting the system and ancilla, which starts in some state ${\ket{0}}_\alpha$, via a unitary operation $U$ and then measuring a state ${\ket{\phi}}_\alpha$ on $\alpha$:
\begin{equation} \label{eqn:kraus_dilation}
	K \rho K^\dagger = {\bra{\phi}}_\alpha U(\rho_S \ox {\ketbra{0}{0}}_\alpha) U^\dagger {\ket{\phi}}_\alpha.
\end{equation}

Suppose that the unitary $U$ is required to be incoherent with respect to $S$, while any operation is allowed on $\alpha$. This means that, for any pure state ${\ket{\psi}}_\alpha$ of the ancilla, $U$ must have the action
\begin{equation} \label{eqn:incoherent_unitary_1}
	U {\ket{i}}_S {\ket{\psi}}_\alpha = {\ket{f(i)}}_S {\ket{\psi'(i)}}_\alpha \quad \forall i,
\end{equation}
where ${\ket{\psi'(i)}}_\alpha$ can be arbitrary. Such a unitary can never create a superposition of basis states of $S$ when viewed at a global level.

Bearing in mind that $f$ must be invertible, it follows that the most general such unitary can be written as
\begin{equation} \label{eqn:incoherent_unitary}
	U = \sum_{i=0}^{d_s-1} {\ketbra{\pi(i)}{i}}_S \ox (U_i)_\alpha,
\end{equation}
where $d_S$ is the dimension of the system, $\pi$ is a permutation acting on $\{0,1,\dots,d_S-1\}$ and the $U_i$ are arbitrary unitary operators on $\alpha$.

\begin{figure}[h]
	\centering
	\includegraphics[scale=.95]{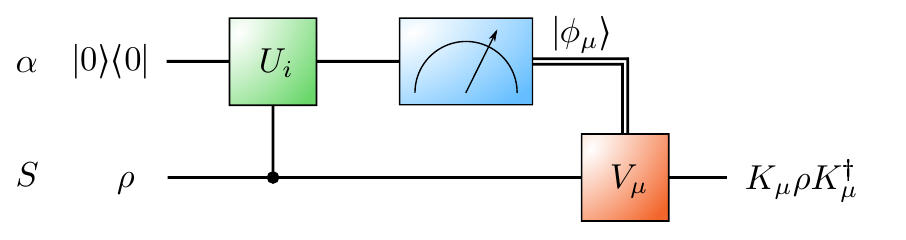}
	\caption{Strictly incoherent operations on system $S$ constructed by an incoherent unitary interaction with an ancilla $\alpha$. An operation is split into a controlled unitary $\sum_i {\ketbra{i}{i}}_S \ox (U_i)_\alpha$ followed by a measurement on the ancilla in a basis $\ket{\phi_\mu}$ and an incoherent unitary $V_\mu$ on the system conditioned on the measurement outcome $\mu$.}
	\label{fig:dilation}
\end{figure}

We allow the ancilla to be measured in any basis $\{{\ket{\phi_\mu}}_\alpha\}$ and any incoherent unitary $V_\mu = \sum_i e^{i \theta^\mu_i} \ketbra{\pi_\mu(i)}{i}$ to be applied to $S$ conditioned on the result ${\ket{\phi_\mu}}_\alpha$ -- see Fig.\ \ref{fig:dilation} for an illustration. Given the freedom in $V_\mu$, we can assume the permutation $\pi$ in $U$ to be trivial. With this in place, we have the following (see Appendix \ref{app:dilation} for the proof):

\begin{res} \label{res:dilation}
	An operation on a system $S$ is SI if and only if it can be constructed from the following elementary processes using an ancilla $\alpha$:
	\begin{enumerate}
		\item Unitary operations on $\alpha$ controlled by the incoherent basis of $S$: $\sum_i {\ketbra{i}{i}}_S \ox (U_i)_\alpha$;
		\item Measurements on $\alpha$ in any basis;
		\item Incoherent unitary operations on $S$: $\sum_i e^{i \theta_i} \ketbra{\pi(i)}{i}$, allowed to be conditioned on the measurement outcome.
	\end{enumerate}
\end{res}

One interpretation of this is that, when viewed at a global level with an ancilla, processes that measure in a coherent basis are as difficult to perform as processes that create coherence. Therefore this may be seen as an another operational motivation for the SI operations.

In the interferometer picture described in the previous section, the branch degree of freedom plays the role of the system $S$, while the internal state of the particle plays the role of the ancilla $\alpha$. The controlled unitary $U$ is equivalent to a set of path-dependent unitary operations, and the permutations $V_\mu$ are represented by reordering branches.

\section{Monotones}
We recall the criteria required for a quantity $M$ to be a measure of coherence, as proposed in Ref.\ \cite{baumgratz2014quantifying}: $M(\rho) \geq 0$, with equality if and only if $\rho$ is incoherent; $M$ cannot increase under incoherent operations. The latter condition is known as monotonicity, and can be given in two different forms (using language borrowed from Ref.\ \cite{gour2008resource}):

\begin{defn} \label{def:monotones}
	Under a trace-preserving (i.e., deterministic) operation $\mc{E} = \sum_\mu \mc{E}^\mu$ with $p_\mu \rho^\mu := \mc{E}^\mu(\rho)$ and $\tr \rho^\mu=1$, a real-valued quantity $M(\rho) \geq 0$ is said to be
	\begin{itemize}
		\item a \emph{deterministic monotone} if $M(\mc{E}(\rho)) \leq M(\rho)$ for all states $\rho$;
		\item an \emph{ensemble monotone} if $\sum_\mu p_\mu M(\rho^\mu) \leq M(\rho)$ for all states $\rho$.
	\end{itemize}
\end{defn}

The former is generally considered to be more fundamental than the latter. Another useful property is convexity, such that $M(\sum_i p_i \rho_i) \leq \sum_i p_i M(\rho_i)$ for any set of states $\rho_i$ and probabilities $p_i$.

A measure which satisfies all of the above criteria is the relative entropy of coherence \cite{baumgratz2014quantifying}:
\begin{align} \label{eqn:rel_ent}
C(\rho) & := \min_{\sigma \text{ incoherent}} S(\rho || \sigma) \nonumber \\
& = \min_{\sigma \text{ incoherent}} -S(\rho) - \tr(\rho \log \sigma) \nonumber \\
& = S(\Phi(\rho)) - S(\rho),
\end{align}
where $S(\rho) = -\tr(\rho \log \rho)$ is the von Neumann entropy. The second line re-expresses the general definition of the quantum relative entropy, and the third is a simplification in the present case.

More generally, a whole class of measures can be constructed from distance measures $D$ with two properties: $D(\rho,\sigma)=0$ if and only if $\rho=\sigma$; $D$ is contractive under trace-preserving quantum operations $\mc{E}$, meaning that $D(\mc{E}(\rho),\mc{E}(\sigma)) \leq D(\rho,\sigma)$. The measure of coherence associated with $D$ is $C_D(\rho) := \min_{\sigma \text{ incoherent}} D(\rho,\sigma)$ \cite{baumgratz2014quantifying}.

Since the SI operations are a subset of the incoherent operations, any coherence measure defined with respect to incoherent operations is a monotone under SI operations. Conversely, we present a set of measures that are deterministic monotones under SI but not necessarily all incoherent operations:
\begin{equation} \label{eqn:dephased_distance}
C'_D(\rho) := D(\rho, \Phi(\rho)).
\end{equation}
It follows from $[\mc{E},\Phi]=0$ that measures of the form (\ref{eqn:dephased_distance}) are deterministic monotones under SI operations \footnote{This was shown in Ref.\ \cite{streltsov2015genuine} for the special case of GI operations, as defined in Section \ref{sec:comparison}.}:
\begin{align}
C'_D(\mc{E}(\rho)) &= D(\mc{E}(\rho), \Phi \circ \mc{E}(\rho)) \nonumber \\
&= D(\mc{E}(\rho), \mc{E} \circ \Phi(\rho)) \nonumber \\
& \leq D(\rho, \Phi(\rho)) \nonumber \\
&= C'_D(\rho).
\end{align}

$C'_D$ in general differs from $C_D$, as there are measures for which the closest incoherent state to $\rho$ is not necessarily $\Phi(\rho)$. For instance this is known to be the case for the fidelity of coherence \cite{shao2015fidelity}
\begin{equation}
C_f(\rho) := \min_{\sigma \text{ incoherent}} 1 - F(\rho,\sigma),
\end{equation}
where $F(\rho,\sigma) := \tr \sqrt{\sqrt{\rho}\sigma\sqrt{\rho}}$. (This is also true for the geometric measure of coherence \cite{streltsov2015measuring} where $F(\rho,\sigma)$ is replaced by $F(\rho,\sigma)^2$.) Therefore $C'_f(\rho) := 1 - F(\rho,\Phi(\rho))$ may be an additional monotone obtained by restricting to SI operations. However, we have not been able to find any examples where monotonicity under incoherent operations is violated.

\section{Comparison with other work on coherence}
\label{sec:comparison}

\subsection{Coherent transport}
Levi and Mintert \cite{levi2014quantitative} have described a set of processes that should be considered classical in transport mechanisms. These are such that any localised excitation must remain local, and the number of excitations cannot change. They showed that these operations can be generated by combinations of two elementary types of processes. Restricting to the single exciting subspace and denoting the state of a particle localised at site $i$ as $\ket{i}$, these processes are:
\begin{enumerate}
	\item ``Modification of phase coherence": Kraus operators $A_i = u_1 \ketbra{i}{i} + u_2 \sum_{j\neq i} \ketbra{j}{j}$;
	\item ``Incoherent hopping": Kraus operators $B_{ji} = \ketbra{j}{i}$.
\end{enumerate}
It is clear that these form a strict subset of the SI operations. For instance, an SI operation could take $\ket{1} + \ket{2} \to \ket{2} + \ket{3}$, but this would be impossible under the above processes.

\subsection{Asymmetry}
The resource theory of quantum reference frames, or asymmetry \cite{bartlett2007reference,gour2008resource}, has also been suggested as a framework for coherence \cite{marvian2014modes}. In this approach, one considers a coherent state to be a phase reference for the phase conjugate to some preferred observable $A$. A coherent state is then asymmetric with respect to the transformations $T_\theta: \rho \to e^{-i \theta A} \rho e^{i \theta A},\, \theta \in \mathbb{R}$. The incoherent states coincide with those considered here, being mixtures of the eigenstates of $A$.

The free (or \emph{covariant}) operations are defined differently. They are the operations which are possible to perform without access to a phase reference, and satisfy $T_\theta(\mc{E}(\rho)) = \mc{E}(T_\theta(\rho))$ for all $\theta$ and $\rho$. A covariant operation always has a Kraus operator representation of the form $K_{\delta} = \sum_{i,j:\, a_i - a_j = \delta}  c_{i,j} \ketbra{i}{j}$, where $\ket{i}$ and $a_i$ are the eigenstates and eigenvalues of $A = \sum_i a_i \ketbra{i}{i}$ \cite{gour2008resource}. Such a $K_{\delta}$ connects basis states whose values of $A$ differ by a fixed amount $\delta$. When $A$ has no degeneracy, the covariant operations are a subset of the incoherent operations \cite{yadin2016general,marvian2015quantum} -- moreover, we see that they are a strict subset of the SI operations. The only extra restriction is on the allowed reshuffling of basis states. A covariant operation associated with $K_\delta$ may only shift states rigidly up or down in the spectrum of $A$. This means that coherence associated with off-diagonal elements $\ketbra{i}{j}$ for different values of $a_i-a_j$ constitute separate resources that cannot be interconverted.

Similarly to our physical picture in Result \ref{res:dilation}, the resource theory of asymmetry also has a unitary dilation theorem \cite{keyl1999optimal,marvian2012symmetry}. In this, the global unitary is covariant with respect to phase conjugate to a global observable $A \ox I + I \ox B$, where $B$ is some suitably chosen observable on the environment, and the initial state of the environment is incoherent in the basis of $B$. For example, consider a situation where the total number of particles is conserved. Then an environment with no coherence in the number basis is given for free, and after interaction the resulting operation on the system is covariant.

An environment containing coherence acts as a phase reference and thus constitutes a resource for overcoming the global symmetry \cite{marvian2008building,ahmadi2013wigner,navascues2014energy,aaberg2014catalytic}. However, in the SI picture, any operation is allowed on the environment for free, so its coherence is irrelevant. Furthermore, global symmetric unitaries are able to make use of degeneracy in the global observable. For example, a superposition of number states $\alpha \ket{0}\ket{1} + \beta \ket{1}\ket{0}$ can be created for free in the resource theory of asymmetry. Thus more operations become available when systems are combined. As defined here, SI operations allow for no such degeneracy.

\subsection{Genuinely incoherent operations}
\begin{figure}[h]
	\centering
	\includegraphics[scale=0.7]{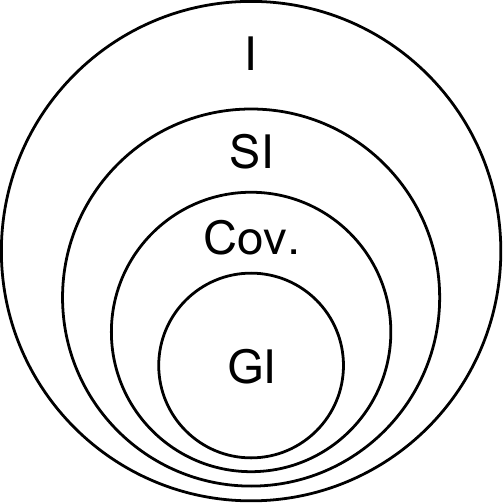}
	\caption{Hierarchy of classes of incoherent operations: incoherent (I), strictly incoherent (SI), covariant (Cov.) and genuinely incoherent (GI).}
	\label{fig:hierarchy}
\end{figure}

Streltsov \cite{streltsov2015genuine} recently defined the subset of the incoherent operations for which every possible set of Kraus operators is incoherent. They are also such that every incoherent state is unchanged (in the trace-preserving case). These operations can be recovered from our dilation picture in Result \ref{res:dilation} by removing the conditional operations $V_\mu$:

\begin{defn}
	An SI operation which can be constructed from a global incoherent process as in Result \ref{res:dilation}, without the use of the conditional operations $V_\mu$, is said to be \emph{genuinely incoherent} (GI). This has a set of Kraus operators of the form $K_\mu = \sum_i c^\mu_i \ketbra{i}{i}$.
\end{defn}
Evidently these form a strict subset of the SI operations. To summarise (see Fig.\ \ref{fig:hierarchy}), we have
\begin{align} \label{eqn:hierarchy}
\text{incoherent operations} \supset \text{strictly incoherent operations} \nonumber \\
\supset \text{covariant operations} 
\supset \text{genuinely incoherent operations.}
\end{align}

\section{Application to quantum correlations}
In the following sections, we demonstrate the role played by SI operations in characterising quantum correlations. Firstly, we define a measure of quantumness of correlations with respect to the incoherent basis, a basis-dependent version of quantum discord. We find the conditions for it to vanish, and to be a monotone under local SI operations.

The form of the dilation picture in Result \ref{res:dilation} plays an important role here in a subtle distinction between deterministic and ensemble monotonicity. We find that the basis-dependent discord is an ensemble monotone under local SI operations, when the ensemble is selected by the measurement outcomes in the dilation. However, it is not generally a deterministic monotone, which can be traced back to a loss of classical correlations by `forgetting' the measurement outcome.

Next, we interpret the vanishing condition as saying that a state can be recovered after local dephasing. This leads to a new class of measures quantifying approximate recoverability, which have the same behaviour as the discord measure under SI operations. Finally, we provide a connection with the basis-independent discord by examining the set of operations which are SI in every basis.

\subsection{Definitions}
The total correlations in a bipartite quantum state $\rho_{AB}$ may be quantified by the mutual information ${I(A:B)_\rho} = S(\rho_A) + S(\rho_B) - S(\rho_{AB})$. A standard measure of ``classical correlations" is given by the largest mutual information that can be obtained when one side undergoes any (POVM) measurement $\mc{M}$ (though one could define a symmetric case where both sides are measured): $\max_{\mc{M}_A} {I(A:B)_{\mc{M}_A(\rho_{AB})}}$. The quantum discord is defined as the difference between the total correlations and the classical correlations \cite{modi2012classical}.

In the context of coherence, however, the word ``classical" takes on a particular meaning: we can consider the classical correlations to be the mutual information shared between a classical observer at $A$ and another observer at $B$:
\begin{equation} \label{eqn:classical_correlations}
	J(B|A)_\rho := I(A:B)_{\Phi_A(\rho)},
\end{equation}
where $\Phi_A(\rho_{AB}) = \sum_i (\ketbra{i}{i} \ox I) \rho_{AB} (\ketbra{i}{i} \ox I)$.

The discord with respect to this fixed measurement then follows as the difference
\begin{equation} \label{eqn:discord}
	\delta(B|A)_\rho := I(A:B)_\rho - J(B|A)_\rho.
\end{equation}
We denote the standard discord, which includes a minimisation over all measurements, by $\dmin$. It should be noted that one of the original papers on discord, Ref.\ \cite{ollivier2001quantum}, initially motivated the measurement-dependent case before discussing minimisation.

One of the usual requirements of a measure of (quantum) correlations is that it be invariant under local changes of basis -- $\dmin$ satisfies this, while $\delta$ does not. However, once we understand $J$ as a measure of classical correlations with respect to the preferred incoherent basis, $\delta$ can be seen as a basis-dependent measure of quantumness of correlations, in the same way that coherence is a basis-dependent measure of quantumness in single systems.

Following Ref.\ \cite{ma2015converting}, we note that $\delta$ can be written as the difference between measures of coherence in the global and local states:
\begin{equation} \label{eqn:discord_coherence}
	\delta(B|A)_\rho = C(B|A)_\rho - C(A)_\rho.
\end{equation}
Here, $C(A)_\rho := C(\rho_A)$ is the relative entropy of coherence in the reduced state $\rho_A$. $C(B|A)_\rho$ is defined similarly  as $\min_{\sigma_{AB} \in IQ} S(\rho_{AB}||\sigma_{AB})$, where the minimisation is over \emph{incoherent-quantum} (IQ) states \cite{chitambar2015assisted} of the form $\sigma_{AB} = \sum_i p_i \ketbra{i}{i} \ox \rho_{B|i}$.

\subsection{Vanishing basis-dependent discord}
The standard discord $\dmin(B|A)$ vanishes if and only if the state is classical-quantum (CQ) \cite{modi2012classical}, meaning that there is a basis $\{\ket{\psi_a}\}$ of $A$ such that
\begin{equation} \label{eqn:cq}
	\rho_{AB} = \sum_a \lambda_a \ketbra{\psi_a}{\psi_a} \ox \rho_{B|a}.
\end{equation}
In our basis-dependent case, we have
\begin{res} \label{res:zero_discord}
	$\delta(B|A)_\rho = 0$ if and only if there exists a decomposition $\rho_{AB} = \sum_\alpha p_\alpha \rho_A^\alpha \ox \rho_B^\alpha$ such that all $\rho_A^\alpha$ are perfectly distinguishable by measurements in the incoherent basis.
\end{res}

See Appendix \ref{app:zero_discord} for the proof. These zero-$\delta$ states may be coherent, but each $\rho_A^\alpha$ component must have support spanned by a different subset of the incoherent basis states. For example, the following qutrit-qubit state has $\delta=0$:
\begin{equation} \label{eqn:ic_example}
	\rho_{AB} = \frac{1}{2} ( \ketbra{+_{01}}{+_{01}} \ox \ketbra{0}{0} + \ketbra{2}{2} \ox \ketbra{1}{1} ),
\end{equation}
where $\ket{\pm_{ij}} := (\ket{i} \pm \ket{j})/\sqrt{2}$.

Since $\dmin \leq \delta$, the states with vanishing $\delta$ must be a subset of the CQ states (one can explicitly verify that the above example is of the form (\ref{eqn:cq}) by taking $\ket{\psi_0}=\ket{+_{01}},\ \ket{\psi_1}=\ket{-_{01}},\ \ket{\psi_2}=\ket{2},\ \rho_0 = \ketbra{0}{0}$ and $\rho_1 = \ketbra{1}{1}$). As (\ref{eqn:ic_example}) shows, contrary to what one might guess, the basis-dependent discord does not vanish \emph{only} for CQ states (\ref{eqn:cq}) where the terms $\ket{\psi_a}$ are just the incoherent basis states. These are in fact just IQ states, characterised by vanishing $C(B|A)$ (and also by vanishing distillable coherence under local incoherent-quantum operations and classical communication \cite{chitambar2015assisted}). The remaining states have $C(B|A) = C(A) > 0$.

Note that there is an error in Ref.\ \cite{ollivier2001quantum} (equation (16)), stating (in our terminology) that only IQ states have $\delta=0$.

In addition to the IQ states, the set of zero-$\delta$ states also contains all product states. In the case where subsystem $A$ is a qubit (i.e., has dimension 2), the IQ and product states are actually the only possibilities. More complex cases such as (\ref{eqn:ic_example}) emerge only in higher dimensions.

\begin{figure}[h]
	\centering
	\includegraphics[scale=.95]{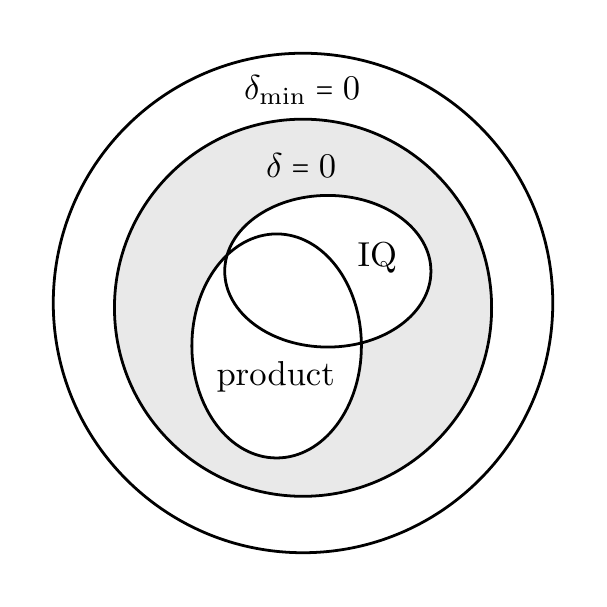}
	\caption{Schematic illustration of the sets for which the following quantities vanish: $\dmin(B|A)$ (CQ states), $\delta(B|A)$ (Result \ref{res:zero_discord}), $C(B|A)$ (IQ states) and $I(A:B)$ (product states). When $A$ is a qubit, the shaded region is empty.}
	\label{fig:correlations}
\end{figure}

See Fig.\ \ref{fig:correlations} for a depiction of the relations between these sets of states. We discuss the operational significance of the form of the zero-$\delta$ states later.

\subsection{Behaviour under local operations}
We now investigate the behaviour of the classical and quantum correlations under local incoherent and SI operations. It is well known \cite{vedral2002role} that the mutual information ${I(A:B)}$ is both a deterministic and an ensemble monotone under local operations on either $A$ or $B$.

One may expect the measure of classical correlations $J(B|A)$ to be a monotone under some suitable set of classical local operations. It can increase under general local incoherent operations -- this was already noted in Ref. \cite{chitambar2015relating}. In fact, we have the following (with the proof in Appendix \ref{app:classical_monotone}):

\begin{res} \label{res:classical_monotone}
	The SI set is the largest subset of incoherent operations such that the measure of classical correlations $J(B|A)$ is an ensemble monotone under operations on subsystem $A$.
\end{res}

This is rather intuitive, given our earlier characterisation of SI operations -- they can never use coherence present in the state to create correlations associated with probabilities in the incoherent basis. It is also worth noting that $J(B|A)$ is a monotone under arbitrary operations on $B$ -- the proof is the same, using the fact that local operations on $B$ commute with dephasing on $A$.

Our main result about the behaviour of $\delta(B|A)$ under SI operations is the following:

\begin{res} \label{res:discord_monotone}
	$\delta(B|A)$ is an ensemble monotone under SI operations on subsystem A, where each outcome is selected by the measurement of the ancilla in the dilation of Result \ref{res:dilation}.
\end{res}

\begin{figure}[h]
	\centering
	\includegraphics[scale=.95]{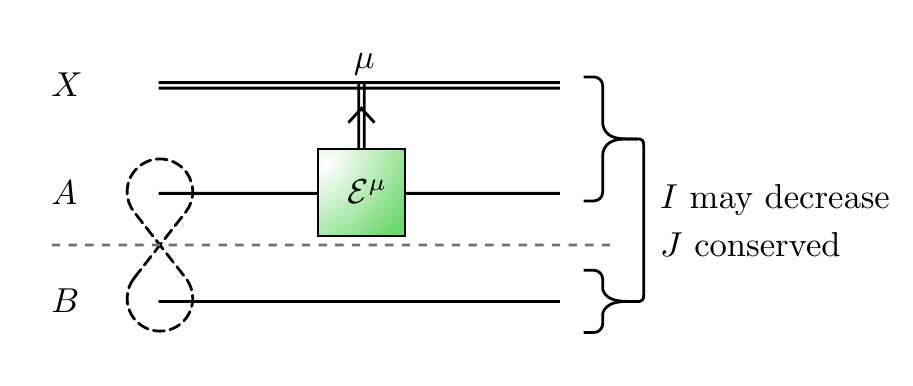}
	\caption{Circuit representation of a local SI operation on a bipartite system $AB$. When an outcome $\mc{E}^\mu$ is obtained by selecting a result $\mu$ from the ancilla measurement in the dilation (Result \ref{res:dilation}), $\mu$ is recorded in a classical memory $X$. Retaining $X$, the total correlations $I(AX:B)$ cannot increase but the classical correlations $J(B|AX)$ are conserved.}
	\label{fig:discord}
\end{figure}

$\delta(B|A)$ is easily seen to be a deterministic monotone under GI operations on $A$, since they leave the incoherent part of the state unchanged -- so $J(B|A)$ is unchanged, while $I(A:B)$ cannot increase. The proof of Result \ref{res:discord_monotone} (see Appendix \ref{app:discord_monotone}) makes use of the fact that all the classical correlations can be recovered when we have access to a classical memory $X$ which records the outcome. This is depicted in Fig.\ \ref{fig:discord}. The main idea is that, as long as party $A$ keeps the memory $X$, the total correlations cannot increase, while the classical correlations are unchanged:
\begin{align}
	I(AX:B)_{\rho'} \leq I(A:B)_\rho, \nonumber \\
	J(B|AX)_{\rho'} = J(B|A)_\rho,
\end{align}
where $\rho$ and $\rho'$ are the input and output states respectively.

There is an important point to be made here, which is that $\delta(B|A)$ is not a \emph{deterministic} monotone under all SI operations. The increase can only come about when mixing different permutations, i.e. when conditional operations $V_\mu$ are used in the global picture. For an explicit example, consider a system starting in the state (\ref{eqn:ic_example}) with $\delta=0$. Applying to subsystem $A$ the operation which is an equal mixture of the identity and the permutation interchanging $\ket{1}$ and $\ket{2}$ results in
\begin{align}
	\rho'_{AB} = \frac{1}{4} \left[ (\ketbra{+_{01}}{+_{01}} + \ketbra{+_{02}}{+_{02}}) \ox \ketbra{0}{0} \right. \nonumber \\
	\quad + \left. (\ketbra{2}{2} + \ketbra{1}{1}) \ox \ketbra{1}{1} \right],
\end{align}
with $\delta(B|A) = \frac{3}{4}\log 3 \approx 0.189 > 0$.

In the picture of Fig.\ \ref{fig:discord}, this increase occurs when tracing out the memory $X$, resulting in the loss of classical information. Thus the classical correlations can decrease, which is necessary for $\delta$ to increase.

We can also view the operations as converting quantumness (as measured by coherence) into quantumness of correlations; this cannot happen when the initial state is incoherent. Such a conversion also clearly requires nonzero total correlations initially. As discussed after Result \ref{res:zero_discord}, when $A$ is a qubit, $\delta=0$ only for IQ and product states, so this cannot happen.

We summarise the observations in the following way:

\begin{res} \label{res:coherence_to_discord}
	$\delta(B|A)$ is a deterministic (and ensemble) monotone under local GI operations. However, a local SI operation can in general create nonzero $\delta(B|A)$ from a state with $\delta(B|A)=0$ but nonzero coherence and correlations.
\end{res}
Stated differently, given a state $\rho$ with $\delta(B|A)_\rho = 0$ but both $C(A)_\rho >0$ and $I(A:B)_\rho>0$, a local GI operation $\mc{G}_A$ always results in $\delta(B|A)_{\mc{G}_A(\rho)} = 0$, but a local SI operation $\mc{E}_A$ in general can give $\delta(B|A)_{\mc{E}_A(\rho)} > 0$.

From (\ref{eqn:discord_coherence}), as in Ref.\ \cite{ma2015converting}, we can bound the increase in $\delta$ by the change in local coherence:
\begin{align}
	\delta(B|A)_{\mc{E}_A(\rho)} - \delta(B|A)_\rho & = C(B|A)_{\mc{E}_A(\rho)} - C(A)_{\mc{E}_A(\rho)} \nonumber \\
		& \quad - C(B|A)_\rho + C(A)_\rho \nonumber \\
		& \leq C(A)_\rho - C(A)_{\mc{E}_A(\rho)},
\end{align}
using the fact that $C(B|A)$ cannot increase under incoherent operations. Hence local coherence must be consumed in order to increase $\delta$.

\subsection{Recovery from local incoherent measurement}
We can gain further insight about the meaning of $\delta$ by interpreting the condition for it to vanish. A state has $\delta=0$ if and only if it can be perfectly locally recovered after a local incoherent measurement -- i.e., if there is some local operation $\mc{R}_A$ such that $\mc{R}_A \circ \Phi_A(\rho_{AB}) = \rho_{AB}$. The demonstration of this fact can be seen by examining our proof of Result \ref{res:zero_discord}, in particular the statement about recovery channels using a theorem by Petz \cite{petz2003monotonicity}. Intuitively, all of the nonlocal information in such a state is stored independently of the coherence.

For $\delta>0$, it is natural to ask whether the degree of recoverability is related to basis-dependent discord. This gives a correspondence with the approach of Ref.\ \cite{seshadreesan2015fidelity} to discord, where the state is disturbed by a measurement on $A$ (in particular, an entanglement-breaking channel \cite{horodecki2003entanglement}), after which the party in control of $A$ tries to recover the original state by local operations. In our case, we have a local incoherent measurement $\Phi_A$ as a specific entanglement-breaking channel.

Given a suitable distance measure $D$, we can quantify the ability to recover the state by
\begin{equation}
	\Delta_D(B|A)_\rho := \min_{\mc{R}_A} D(\rho_{AB}, \mc{R}_A \circ \Phi_A(\rho_{AB})),
\end{equation}
where $\mc{R}_A$ can be any operation on $A$. Firstly, note that $\Delta_D(B|A)=0 \Leftrightarrow \delta(B|A)=0$. Furthermore, $\Delta_D(B|A)$ shares similar monotonicity properties with $\delta(B|A)$. Instead of proving it to be an ensemble monotone, we have monotonicity when the memory $X$ of the outcome is retained. Given that the party in control of $A$ has access to $X$, we find that

\begin{res} \label{res:recovery_monotone}
	For any local SI operation $\mc{E} = \sum_\mu \mc{E}^\mu$ taking $\rho \to \rho'$ while keeping a memory $X$ of the outcome $\mu$, $\Delta_D(B|XA)_{\rho'} \leq \Delta_D(B|A)_\rho$.
\end{res}

See Appendix \ref{app:recovery:monotone} for the proof. We assume that the distance $D$ is contractive under trace-preserving quantum operations.

To summarise, $\Delta_D$ and $\delta$ vanish for the same set of states and are monotonic in the same way under SI operations. Hence one could justify $\Delta_D$ as providing an additional set of basis-dependent discord measures.

\subsection{Implications for basis-independent quantum discord}
We now discuss the possible implications for the theory of basis-independent discord relating to $\dmin$. No resource theory of discord currently exists -- the free states would have to be the CQ states, but the free operations are unknown. What is known, however, is the full set of local operations (on subsystem $A$) which cannot create discord \cite{streltsov2011behavior,yu2011quantum,hu2012necessary,guo2013necessary}. The correct free local operations must be contained in this set. Known as commutativity-preserving operations, these are such that any two commuting states remain commuting: with $\rho \to \rho'$ and $\sigma \to \sigma',\ [\rho,\sigma]=0 \Rightarrow [\rho',\sigma']=0$.

Commutativity-preserving operations can be separated into two main classes. Those in the first class are termed semiclassical: such an operation always outputs a state which is incoherent in some fixed basis. These always completely destroy discord, and hence are of limited interest.

The second class is more subtle and depends on the dimension $d$ of subsystem $A$. For $d=2$ it consists of all unital channels, i.e., satisfying $\mc{E}(I) = I$. For $d>2$, it contains the so-called isotropic channels of the form
\begin{equation} \label{eqn:isotropic}
	\mc{E}(\rho) = p \Gamma(\rho) + (1-p) \frac{I}{d},
\end{equation}
where $\Gamma$ is either a unitary or antiunitary channel, and $p$ is any real number suitably chosen to ensure that $\mc{E}$ is completely positive.

We now provide a connection between the SI operations and a rather large subset of the commutativity-preserving channels, namely the isotropic channels with unitary $\Gamma: \rho \to U \rho U^\dagger$,
\begin{equation} \label{eqn:isotropic_unitary}
	\mc{E}(\rho) = p U \rho U^\dagger + (1-p) \frac{I}{d}.
\end{equation}
To do so, we first define the set of depolarising channels in dimension $d$ as
\begin{equation}
\mc{D} := \{ \mc{E} : \rho \to p \rho +  (1-p) I/d \;| \; p \in \mathbb{R},\, \mc{E} \text{ is a channel} \}.
\end{equation}

\begin{res} \label{res:every_basis}
	A channel is SI in \emph{every} basis if and only if it is depolarising. As a corollary, a channel $\mc{E}$ is of the unitary-isotropic form (\ref{eqn:isotropic_unitary}) if and only if $\mc{E} = \mc{U} \circ \mc{F}$, where $\mc{F}$ is SI in every basis and $\mc{U}$ is unitary.
	
	The only channel which is GI in every basis is the identity.
\end{res}

See Appendix \ref{app:every_basis} for the proof. The second statement follows from the fact that $U[p \rho + (1-p)I/d]U^\dagger = p U \rho U^\dagger + (1-p)I/d$.

This is the first connection between the free operations of coherence and discord. Our proof also makes explicit use of the SI operations, as opposed to the incoherent ones in the sense of Ref.\ \cite{baumgratz2014quantifying}.

\section{Conclusion}
In summary, we have characterised the operations which can neither create nor use coherence, providing a boundary between processes which make use of quantum resources and those that only use classical resources. We have showed that they are mathematically captured by the class of SI operations, and that they admit an operational interpretation in a simple interferometric scheme. This makes the SI class a strong candidate for the set of free operations in the resource theory of coherence currently under developement.

Widening the view to multipartite systems, we have seen that SI operations also provide a novel connection between coherence and quantum correlations. In this instance, the quantumness of correlations is measured with respect to the preferred basis. We have shown that these correlations cannot increase under local SI operations in which the outcome is recorded in a classical memory. Losing this memory, quantum correlations can be created at the expense of local coherence. Furthermore, we have characterised such correlations via new measures, in terms of the global information lost through the local removal of coherence, which share the same behaviour under SI operations. Thus our results suggest a deeper understanding of quantum correlations via the manipulation of coherence, at the same time giving an additional meaning to SI operations. We have also provided the first connection between the free operations for coherence and those for discord -- even though a resource theory for discord is not yet known. Thus we expect our results to pave the way for such a theory.

This work leads to a number of further lines of research. Winter and Yang \cite{winter2016operational} have discussed the operational significance of the resource theory of coherence in terms of distillation of coherent states and the coherence cost of state formation. They have noted that all of their results hold when only SI operations are used -- except for distillation from mixed states, which remains an open question. In addition, the manipulation of coherence under local SI operations and classical communications is yet to be studied, and may provide more links with entanglement. Similarly, one could study how the basis-dependent discord behaves when classical communication is allowed. Further links between coherence and the potential resource theory of discord remain to be investigated, for instance, whether one can deduce the behaviour of basis-independent discord under local operations which are SI in every basis.

\acknowledgements
We thank Tillman Baumgratz, Andrew Garner, Jayne Thompson, Mark Wilde, Andreas Winter, Raam Uzdin, Gerardo Adesso, Marco Cianciaruso and Thomas Bromley for fruitful discussions. This work was supported by the EPSRC (UK) Grant EP/L01405X/1, the John Templeton Foundation Grant 54914, the National Research Foundation NRF, NRF-Fellowship (Reference No: NRF-NRFF2016-02), the Ministry Education in Singapore Grant and the Academic Research Fund Tier 3 MOE2012-T3-1-009, the National Basic Research Program of China Grant 2011CBA00300, 2011CBA00302 and the National Natural Science Foundation of China Grant 11450110058, 61033001, 6136113600, the Leverhulme Trust, the Oxford Martin School and the Wolfson College, University of Oxford.

\appendix

\section{Proof of Result \ref{res:dilation}} \label{app:dilation}
Let us first show that any operation constructed from the three elementary operations is indeed SI -- we give Kraus operators of the desired form. Assume an operation is performed as in Fig.\ \ref{fig:dilation}. Suppose we select a single outcome $\mu$ by projecting out a state $\ket{\phi_\mu}$ on the ancilla $\alpha$. Using (\ref{eqn:kraus_dilation}) and taking $V_\mu = \sum_i e^{i \theta^\mu_i} \ketbra{\pi_\mu(i)}{i}$, this gives an operation with a single Kraus operator
\begin{equation}
	K_\mu = \sum_i \braXket{\phi_\mu}{U_i}{0} e^{i\theta_i} \ketbra{\pi_\mu(i)}{i},
\end{equation}
which is evidently SI by inspection. A full set of SI Kraus operators is given over all possible measurement outcomes by selecting each basis state $\ket{\phi_\mu}$.

Conversely, let $\{K_\mu = \sum_i c^\mu_i \ketbra{\pi_\mu(i)}{i},\, \mu=0,1,\dots,k-1\}$ be a set of SI Kraus operators. We now need to construct a suitable controlled unitary $U = \sum_i {\ketbra{i}{i}}_S \ox (U_i)_\alpha$, a basis set ${\ket{\phi_\mu}}$ to be measured on $\alpha$, and a set of incoherent unitaries $V_\mu = \sum_i {\ketbra{\pi_\mu(i)}{i}}_S$.

The condition $\sum_\mu K_\mu^\dagger K_\mu \leq I$ is equivalent to $\sum_\mu \abs{c^\mu_i}^2 \leq 1\, \forall i$, since 
\begin{align}
	\sum_\mu K_\mu^\dagger K_\mu &= \sum_{\mu,i,j} (c^\mu_i)^* c^\mu_j \ket{i}\braket{\pi_\mu(i)}{\pi_\mu(j)}\bra{j} \nonumber \\
	&= \sum_{\mu,i} \abs{c^\mu_i}^2 \ketbra{i}{i}.
\end{align}
We may as well assume that $\sum_\mu K_\mu^\dagger K_\mu = I$, since we can always find another $K_k = \sum_i c^k_i \ketbra{i}{i}$ (which is SI) to satisfy this.

Taking the dimension of $\alpha$ to be $k+1$, the normalisation $\sum_\mu \abs{c^\mu_i}^2=1$ then makes it possible to construct a set of unitary operators $U_i$ on $\alpha$ such that $\braXket{\phi_\mu}{U_i}{0} = c^\mu_i$ with some basis set $\{\ket{\phi_\mu}\}$. After applying the controlled unitary, measuring the state $\ket{\phi_\mu}$ and conditionally applying $V_\mu = \sum_i \ketbra{\pi_\mu(i)}{i}$ on $S$, it follows that the resulting state is $K_\mu \rho K_\mu^\dagger$. Thus we have given a dilation which correctly reproduces the Kraus operators.

\section{Proof of Result \ref{res:zero_discord}} \label{app:zero_discord}
We first check explicitly that states of the required form do indeed have zero $\delta(B|A)$. Defining $\sigma_{AB} := \Phi_A(\rho_{AB}),\, \sigma_A^\alpha := \Phi(\rho_A^\alpha)$, we have
\begin{equation}
	C(B|A)_\rho = -S(\rho_{AB}) - \tr(\rho_{AB} \log \sigma_{AB})
\end{equation}
and will deal with the two terms separately. Using the fact that the $\rho_A^\alpha$ have orthogonal support, it is seen that
\begin{equation}
	S(\rho_{AB}) = H(\mathbf{p}) + \sum_\alpha p_\alpha S(\rho_A^\alpha \ox \rho_B^\alpha),
\end{equation}
where $H(\mathbf{p}) = \sum_\alpha -p_\alpha \log p_\alpha$ is the Shannon entropy of the probabilities $p_\alpha$. Similarly, since the supports of $\sigma^\alpha_{A(B)}$ are orthogonal for different $\alpha$, and respectively contain the supports of $\rho^\alpha_{A(B)}$ (note that $\sigma^\alpha_B = \rho^\alpha_B$),
\begin{equation}
	\tr(\rho_{AB} \log \sigma_{AB}) = \sum_\alpha p_\alpha \tr \left[ (\rho_A^\alpha \ox \rho_B^\alpha) \log (p_\alpha \sigma_A^\alpha \ox \rho_B^\alpha) \right] .
\end{equation}
Putting these together, it follows that
\begin{align*}
C(B|A)_\rho &= - \left[ H(\mathbf{p}) + \sum_\alpha p_\alpha S(\rho_A^\alpha \ox \rho_B^\alpha) \right] \\
		& \quad - \sum_\alpha p_\alpha \tr \left[ (\rho_A^\alpha \ox \rho_B^\alpha) \log (p_\alpha \sigma_A^\alpha \ox \rho_B^\alpha) \right] \\
		&= -H(\mathbf{p}) - \sum_\alpha p_\alpha \left( S(\rho_A^\alpha) + S(\rho_B^\alpha) \right. \\
		& \quad \left. + \log p_\alpha + \tr[\rho_A^\alpha \log \sigma_A^\alpha] + \tr[\rho_B^\alpha \log \rho_B^\alpha] \right) \\
		&= \sum_\alpha p_\alpha \left( -\tr[\sigma_A^\alpha \log \sigma_A^\alpha] - S(\rho_A^\alpha) \right) \\
		&= \left[ H(\mathbf{p}) + \sum_\alpha p_\alpha S(\sigma_A^\alpha) \right] - \left[ H(\mathbf{p}) + \sum_\alpha p_\alpha S(\rho_A^\alpha) \right] \\
		&= S(\sigma_A) - S(\rho_A) \\
		&= C(A)_\rho,
\end{align*}
Hence $\delta(B|A)_\rho = 0$ from (\ref{eqn:discord_coherence}).

To prove the converse, we follow a method similar to that used in Theorem 1 of \cite{piani2008no}. From (\ref{eqn:discord}), we see that $\delta(B|A)_\rho = 0 \Leftrightarrow S(\rho_{AB} || \rho_A \ox \rho_B) = S(\sigma_{AB} || \sigma_A \ox \rho_B)$. A theorem by Petz \cite{petz2003monotonicity} says that the relative entropy is unchanged under a quantum channel $\mc{E}$, meaning that $S(\rho || \sigma) = S(\mc{E}(\rho) || \mc{E}(\sigma))$, if and only if there is a recovery channel $\mc{R}$ satisfying $\mc{R} \circ \mc{E} (\rho) = \rho,\, \mc{R} \circ \mc{E} (\sigma) = \sigma$. Moreover, there is an explicit formula for the recovery channel: 
\begin{equation}
	\mc{R}(X) = \sigma^{1/2} \mc{E}^\dagger ( \mc{E}(\sigma)^{-1/2} X \mc{E}(\sigma)^{-1/2}) \sigma^{1/2}.
\end{equation}

In our case, $\mc{E}$ is the dephasing channel $\Phi_A = \Phi_A^\dagger$. By writing
\begin{equation}
	\sigma_A = \sum_i p_i \ketbra{i}{i} \ox \rho_{B,i},
\end{equation}
the recovery condition for $\rho_{AB}$ says that
\begin{align} \label{eqn:recovered_state}
	\rho_{AB} &= \mc{R}(\sigma_{AB}) \nonumber \\
		&= \sum_i p_i \mc{R}\left( \ketbra{i}{i} \ox \rho_{B,i} \right) \nonumber  \\
		&= \sum_i p_i (\rho_A \ox \rho_B)^{1/2} \nonumber  \\ 
		& \quad \Phi_A \left[ (\sigma_A \ox \rho_B)^{-1/2} (\ketbra{i}{i} \ox \rho_{B,i}) (\sigma_A \ox \rho_B)^{-1/2} \right] \nonumber  \\ 
		& \quad (\rho_A \ox \rho_B)^{1/2} \nonumber  \\
		&= \sum_i \left(\rho_A^{1/2} \ketbra{i}{i} \rho_A^{1/2} \right) \ox \rho_{B,i},
\end{align}
after some simple manipulation.		

Since $\delta(B|A)=0 \Rightarrow \dmin(B|A)=0$, (\ref{eqn:cq}) must apply, i.e.,
\begin{equation}
	\rho_{AB} = \sum_a \lambda_a \ketbra{\psi_a}{\psi_a} \ox \rho_{B|a}.
\end{equation}
By setting this equal to (\ref{eqn:recovered_state}) and pre- and post-multiplying by $\rho_A^{-1/2} \ox I$, we obtain
\begin{equation} \label{eqn:recovered_simplified}
	\sum_i \ketbra{i}{i} \ox \rho_{B,i} = \sum_a \ketbra{\psi_a}{\psi_a} \ox \rho_{B|a},
\end{equation}
Note that we have defined
\begin{equation}
	\rho_A^{-1/2} := \sum_{a:\, \lambda_a \neq 0} \lambda_a^{-1/2} \ketbra{\psi_a}{\psi_a},
\end{equation}
where $\lambda_a$ and $\ket{\psi_a}$ are the eigenvalues and eigenstates of $\rho_A$ respectively. Therefore in (\ref{eqn:recovered_simplified}) we exclude terms on either side that are not in the support of $\rho_A$.

After taking the inner product $\braXket{i}{(\;)}{\psi_a}$ in (\ref{eqn:recovered_simplified}), we have $\braket{i}{\psi_a} \rho_{B,i} = \braket{i}{\psi_a} \rho_{B|a} \; \forall a, i$. Hence either $\braket{i}{\psi_a} = 0$ or $\rho_{B,i} = \rho_{B|a}$.

To arrive at the claimed result, we introduce the concept of the coherence-support of a state as the set of incoherent basis vectors that have nonzero overlap with the state. It is clear that two states are perfectly distinguishable by measurements in the incoherent basis if and only if they have disjoint coherence-support. If two eigenstates $\ket{\psi_a},\ket{\psi_b}$ have intersecting coherence-support, then $\exists i : \; \rho_{B|a} = \rho_{B|b} = \rho_{B,i}$. By grouping together different terms in (\ref{eqn:cq}) containing $\ket{\psi_a}$ with the same associated $\rho_{B|a}$ into a single $\rho_A^\alpha$, the proof is finished.

\section{Proof of Result \ref{res:classical_monotone}} \label{app:classical_monotone}
We assume throughout this proof that $\mc{E}$ is an incoherent operation. First we show that $J(B|A)$ is an ensemble monotone under SI operations. Let $\mc{E}$ be SI, then $\mc{E} = \sum_\mu \mc{E}^\mu$ with $[\mc{E}^\mu,\Phi]=0$. If we consider the channel $\mc{F}(\rho) = \sum_\mu \ketbra{\mu}{\mu} \ox \mc{E}_A^\mu(\rho)$ which performs this operation while retaining a memory of the outcomes, it follows that
\begin{align}
	J(B|A)_{\rho} & = I(A:B)_{\Phi_A(\rho)} \nonumber \\
		& = S(\Phi_A(\rho_{AB}) || \Phi(\rho_A) \ox \rho_B) \nonumber \\
		& \geq S(\mc{F}_A \circ \Phi_A(\rho_{AB}) || \mc{F} \circ \Phi(\rho_A) \ox \rho_B) \nonumber \\
		& = S(\Phi_A \circ \mc{F}_A(\rho_{AB}) || \Phi \circ \mc{F}(\rho_A) \ox \rho_B) \nonumber \\
		& = S \left(\sum_\mu \ketbra{\mu}{\mu} \ox \Phi_A \circ \mc{E}_A^\mu(\rho_{AB}) || \right. \nonumber \\
		& \qquad \left. \sum_\mu \ketbra{\mu}{\mu} \ox \Phi \circ \mc{E}^\mu(\rho_A) \ox \rho_B \right) \nonumber \\
		& = \sum_\mu p_\mu S \left(\ketbra{\mu}{\mu} \ox \Phi_A( \rho_{AB}^\mu) || \ketbra{\mu}{\mu} \ox \Phi(\rho_A^\mu) \ox \rho_B \right) \nonumber \\
		& = \sum_\mu p_\mu S \left( \Phi_A(\rho_{AB}^\mu) || \Phi(\rho_A^\mu) \ox \rho_B \right) \nonumber \\
		& = \sum_\mu p_\mu J(B|A)_{\rho^\mu},
\end{align}
where we have used the monotonicity of the relative entropy under $\mc{F}$ for the inequality, and the easily checked property that $S(\sum_\mu p_\mu \ketbra{\mu}{\mu} \ox \rho^\mu || \sum_\mu p_\mu \ketbra{\mu}{\mu} \ox \sigma^\mu) = \sum_\mu p_\mu S(\rho^\mu || \sigma^\mu)$.

Conversely, suppose that $[\mc{E}, \Phi] \neq 0$. Then there exist $i,j,k$ with $i \neq j$ such that $\braXket{k}{\mc{E}(\ketbra{i}{j})}{k} \neq 0$. Now let $\sigma_i := \mc{E}(\ketbra{i}{i})$ and similarly for $j$, and $\tau := \mc{E}(\ketbra{i}{j})$.

We construct a state with $J(B|A)=0$ and show that $J(B|A)$ becomes nonzero after applying $\mc{E}$ locally on $A$. The state is
\begin{equation}
	\rho_{AB} := \frac{1}{2} \left( \ketbra{\phi}{\phi} \ox \ketbra{0}{0} + \frac{1}{2} [ \ketbra{i}{i} + \ketbra{j}{j} ] \ox \ketbra{1}{1} \right),
\end{equation}
where $\ket{\phi} := (\ket{i} + e^{i \phi}\ket{j}) / \sqrt{2}$; it is clear that $J(B|A)_\rho = 0$ since
\begin{equation}
	\Phi_A(\rho_{AB}) = \frac{1}{2} \left( \ketbra{i}{i} + \ketbra{j}{j} \right) \ox \frac{1}{2} \left(\ketbra {0}{0} + \ketbra{1}{1} \right).
\end{equation}
Now perform the incoherent operation $\mc{E}_A$, resulting in the state
\begin{align}
	\rho'_{AB} &= \mc{E}_A(\rho_{AB}) \nonumber \\
	& =\frac{1}{2} \left( \frac{1}{2}[\sigma_i + \sigma_j + e^{-i\phi}\tau + e^{i\phi}\tau^\dagger] \ox \ketbra{0}{0} \right.  \nonumber \\ 
	& \qquad \left. + \frac{1}{2}[\sigma_i + \sigma_j] \ox \ketbra{1}{1} \right).
\end{align}
We have $J(B|A)_{\rho'} = 0$ if and only if $\Phi( e^{-i\phi}\tau +e^{i\phi}\tau^\dagger) = 0$. However, by assumption there exists $k$ such that $\braXket{k}{\tau}{k} \neq 0$. Thus we can choose $\phi$ such that $\braXket{k}{e^{-i\phi}\tau + e^{i\phi}\tau^\dagger}{k} \equiv 2( \cos \phi \, \mathrm{Re}[\braXket{k}{\tau}{k}] + \sin \phi \, \mathrm{Im}[\braXket{k}{\tau}{k}]) \neq 0$ and so $J(B|A)_{\rho'} > 0$. Therefore $J(B|A)$ is not monotonic under $\mc{E}$.

\section{Proof of Result \ref{res:discord_monotone}} \label{app:discord_monotone}
As in Appendix \ref{app:classical_monotone} we can associate to any trace-preserving SI operation $\mc{E} = \sum_\mu \mc{E}^\mu$ with $\mc{E}^\mu(\rho) = \sum_\mu K_\mu \rho K_\mu^\dagger$ another operation which keeps a record of the outcome in an additional system $X$: $\mc{F}(\rho_{AB}) = \sum_\mu {\ketbra{\mu}{\mu}}_X \ox \mc{E}^\mu_A(\rho_{AB})$. Since each term satisfies $[\mc{E}^\mu,\Phi]=0$, we have $[\mc{F},\Phi_A]=0$. Hence, defining
\begin{align}
	\sigma_{AB} &:= \Phi_A(\rho_{AB}), \\
	\tilde{\rho}_{XAB} &:= \mc{F}(\rho_{AB}), \\
	\tilde{\sigma}_{XAB} &:= \mc{F}(\sigma_{AB}),
\end{align}
we also find $\tilde{\sigma}_{XAB} = \Phi_A(\tilde{\rho}_{XAB})$.

The final average basis-dependent discord can be written as
\begin{equation}
	\sum_\mu p_\mu \delta(B|A)_{\rho^\mu} = \sum_\mu p_\mu \left[ I(A:B)_{\rho^\mu} - I(A:B)_{\sigma^\mu} \right],
\end{equation}
and it is simple to verify that
\begin{align}
	I(XA:B)_{\tilde{\rho}} &= H(\mathbf{p}) + \sum_\mu p_\mu I(A:B)_{\rho^\mu}, \\
	I(XA:B)_{\tilde{\sigma}} &= H(\mathbf{p}) + \sum_\mu p_\mu I(A:B)_{\sigma^\mu}.
\end{align}
Therefore
\begin{equation} \label{eqn:avg_discord_difference}
	\sum_\mu p_\mu \delta(B|A)_{\rho^\mu} = I(XA:B)_{\tilde{\rho}} - I(XA:B)_{\tilde{\sigma}}.
\end{equation}
Since $\tilde{\rho}_{XAB}$ is obtained from $\rho_{AB}$ by a map local to $A$, it follows that
\begin{equation} \label{eqn:mututal_info_decrease}
	I(XA:B)_{\tilde{\rho}} \leq I(A:B)_\rho.
\end{equation}

On the other hand, $\sigma_A$ and $\sigma_{AB}$ are recoverable from $\tilde{\sigma}_{XA}$ and $\tilde{\sigma}_{XAB}$ respectively. To see this, first write $\sigma_{AB} = \sum_i \rho_{ii} {\ketbra{i}{i}}_A \ox \rho_{B,i}$ and $K_\mu = \sum_i c^\mu_i \ketbra{\pi_\mu(i)}{i}$ so that
\begin{equation}
	\tilde{\sigma}_{XAB} = \sum_{\mu,i} \rho_{ii} \abs{c^\mu_i}^2 {\ketbra{\mu}{\mu}}_X \ox {\ketbra{\pi_\mu(i)}{\pi_\mu(i)}}_A \ox \rho_{B,i}.
\end{equation}
The recovery operation is chosen to be 
\begin{equation}
	\mc{T}: {\ket{\mu}}_X {\ket{\pi_\mu(i)}}_A \to {\ket{i}}_A,
\end{equation}
which is possible since $\pi_\mu$ is a permutation and thus invertible. One can see that $\mc{T}(\tilde{\sigma}_{XAB}) = \sigma_{AB},\, \mc{T}(\tilde{\sigma}_{XA}) = \sigma_A$ as claimed.

Since the mutual information can be written using the relative entropy as 
\begin{align}
	I(A:B)_\sigma &= S(\sigma_{AB} || \sigma_A \ox \sigma_B),\\
	I(XA:B)_{\tilde{\sigma}} &= S(\tilde{\sigma}_{XAB} || \tilde{\sigma}_{XA} \ox \tilde{\sigma}_B),
\end{align}
the existence of such a recovery map implies that
\begin{equation} \label{eqn:mutual_info_const}
	I(XA:B)_{\tilde{\sigma}} = I(A:B)_\sigma.
\end{equation}

Putting (\ref{eqn:mututal_info_decrease}) and (\ref{eqn:mutual_info_const}) into (\ref{eqn:avg_discord_difference}), we find that
\begin{align}
	\sum_\mu p_\mu \delta(B|A)_{\rho^\mu} & \leq I(A:B)_\rho - I(A:B)_\sigma \nonumber \\
		& = \delta(B|A)_\rho.
\end{align}

\section{Proof of Result \ref{res:recovery_monotone}} \label{app:recovery:monotone}
We use the same notation as in Appendix \ref{app:discord_monotone}. In the final state of $XAB$ we have
\begin{align} \label{eqn:recovery_final}
	\Delta_D(B|XA)_{\tilde{\rho}} & = \min_{\mc{R}} D(\tilde{\rho}_{XAB}, \mc{R}(\tilde{\sigma}_{XAB})) \nonumber \\
		& = \min_{\mc{R}} D(\mc{F}(\rho_{AB}), \mc{R} \circ \mc{F}(\sigma_{AB})),
\end{align}
where $\mc{R}$ is any trace-preserving CP map taking $XA \to XA$. The party at $A$ is assumed to be in possession of the memory $X$ and thus can act ``locally" on both $A$ and $X$. Just as in Appendix \ref{app:discord_monotone}, we can find a local recovery map $\mc{T}$ with the action ${\ket{\mu}}_X {\ket{\pi_\mu(i)}}_A \to {\ket{i}}_A$, so that $\mc{T}(\tilde{\sigma}_{XAB}) =  \sigma_{AB}$.

Let us also identify the map $\mc{R}^*$ that optimally recovers the initial state $\rho_{AB}$ after dephasing, satisfying $\Delta_D(B|A)_\rho = D(\rho_{AB}, \mc{R}^*(\sigma_{AB}))$. We then choose a particular recovery channel $\mc{R} = \mc{F} \circ \mc{R}^* \circ \mc{T}$ in the right hand side of (\ref{eqn:recovery_final}). Since this cannot recover the state $\tilde{\rho}_{XAB}$ better than the optimal operation, we have
\begin{align}
	\Delta_D(B|XA)_{\tilde{\rho}} & \leq D(\mc{F}(\rho_{AB}), \mc{R} \circ \mc{F}(\sigma_{AB})) \nonumber \\
		& = D(\mc{F}(\rho_{AB}), \mc{F} \circ \mc{R}^* \circ \mc{T} (\tilde{\sigma}_{XAB})) \nonumber \\
		& = D(\mc{F}(\rho_{AB}), \mc{F} \circ \mc{R}^* (\sigma_{AB})) \nonumber \\
		& \leq D(\rho_{AB}, \mc{R}^*(\sigma_{AB})) \nonumber \\
		& = \Delta_D(B|A)_\rho,
\end{align}
where in the penultimate line we have used the contractivity of the distance $D$ under the map $\mc{F}$.

\section{Proof of Result \ref{res:every_basis}} \label{app:every_basis}
We denote the set of incoherent states in basis $b$ by $I_b$. Similarly, we use $\mc{SI}_b,\, \mc{GI}_b$ and $\mc{MI}_b$ for the sets of SI, GI and maximal incoherent operations with respect to $b$, respectively. The latter set, also known as coherence non-generating, is defined as containing all quantum channels which map $I_b$ to itself \cite{winter2016operational}.

Before proving the main result, we have a number of Lemmas.

\begin{lem}
	$\bigcap_b \mc{GI}_b = \{I\}$ (i.e., the identity operation).
	\proof
	The definition of GI operations can be written as
	\begin{equation}
	\mc{E} \in \mc{GI}_b \Leftrightarrow \forall \rho \in I_b,\, \mc{E}(\rho) = \rho.
	\end{equation}
	For an arbitrary state $\rho$, if we choose $b$ to be its eigenbasis, then trivially $\rho \in I_b$. Hence every $\mc{E} \in \mc{GI}_b$ must satisfy $\mc{E}(\rho) = \rho$.
\end{lem}

\begin{lem} \label{lem:p_range}
	In dimension $d$, the set of allowed values for $p$ in the set $\mc{D}$ is
	\begin{equation}
	\frac{-1}{d^2-1} \leq p \leq 1.
	\end{equation}
	\proof
	We first introduce the entangled states on two copies of the system $\ket{\alpha_{kl}} := \sum_{j=0}^{d-1} \frac{1}{\sqrt{d}} \omega^{jk} \ket{j}\ket{j \oplus l}$, where $k,l \in \{0,1,\dots,d-1\},\ \omega := e^{2\pi i/d}$ and $j \oplus l := j + l \mod d$. One can check that these form an orthonormal basis for the total dimension-$d^2$ Hilbert space.
	
	Recall that a map $\mc{E}$ is completely positive if and only if $0 \leq \sigma := (\mc{E}\ox I)(\ketbra{\alpha_{00}}{\alpha_{00}})$ \cite{nielsen2010quantum}. It is straightforward to see that for $\mc{E} \in \mc{D}$,
	\begin{align}
	\sigma &= p \ketbra{\alpha_{00}}{\alpha_{00}} + \frac{(1-p)}{d^2} I \ox I \nonumber \\
	&= p \ketbra{\alpha_{00}}{\alpha_{00}} + \frac{(1-p)}{d^2} \sum_{k,l} \ketbra{\alpha_{kl}}{\alpha_{kl}} \nonumber \\
	&= \left( \frac{1+(d^2-1)p}{d^2} \right) \ketbra{\alpha_{00}}{\alpha_{00}} + \frac{(1-p)}{d^2} \sum_{(k,l)\neq (0,0)} \ketbra{\alpha_{kl}}{\alpha_{kl}}.
	\end{align}
	This is a spectral decomposition for $\sigma$, and all eigenvalues are nonnegative if and only if $p \in [-1/(d^2-1),1]$.
\end{lem}

\begin{lem} \label{lem:unital_subset}
	$\bigcap_b \mc{MI}_b$ is a subset of the unital channels.
	\proof
	The maximally mixed state $I/d$ is the unique state which is incoherent in every basis. So for $\mc{E} \in \bigcap_b \mc{MI}_b$, $\mc{E}(I/d)$ must be incoherent in every basis, and thus equals $I/d$.
\end{lem}

We proceed to use Lemmas \ref{lem:p_range} and \ref{lem:unital_subset} to prove the main statement of Result \ref{res:every_basis}, that $\bigcap_b \mc{SI}_b = \mc{D}$. We do this by proving the inclusions $\mc{D} \subseteq \bigcap_b \mc{SI}_b$ and $\bigcap_b \mc{SI}_b \subseteq \mc{D}$.

Firstly, to show that $\mc{D} \subset \mc{SI}_b\ \forall b$, we construct a set of SI Kraus operators in an arbitrary basis $b = \{\ket{i}\}_{i=0,\dots,d-1}$ for every operation in $\mc{D}$. For $k,l \in \{0,1,\dots,d-1\}$, define
\begin{align}
	K_{kl} &:= \kappa_{kl} L_{kl}, \\
	L_{kl} &:= \sum_{j=0}^{d-1} \omega^{jk} \ketbra{j \oplus l}{j}, \\
	\kappa_{kl} &:= \begin{cases}
		\frac{\sqrt{1+(d^2-1)p}}{d} & k=l=0, \\
		\frac{\sqrt{1-p}}{d} & \text{otherwise}.
	\end{cases}
\end{align}
The operators $K_{kl}$ are immediately seen to be SI in the chosen basis. By Lemma \ref{lem:p_range}, the coeffecients $\kappa_{kl}$ are real since $p \in [-1/(d^2-1),1]$. We find that $K_{kl}^\dagger K_{kl} = \kappa_{kl}^2 I$ and thus
\begin{equation}
	\sum_{k,l} K_{kl}^\dagger K_{kl} = \left[\frac{1+(d^2-1)p}{d^2} + \frac{(1-p)}{d^2} (d^2-1) \right] I = I,
\end{equation}
as required for a trace-preserving map. Next we verify the action of the Kraus operators on the matrix elements of any state:
\begin{align}
	\sum_{k,l} & K_{kl} \ketbra{j}{j'} K_{kl}^\dagger = K_{00}\ketbra{j}{j'}K_{00}^\dagger \nonumber \\
	& \qquad \qquad \qquad + \sum_{(k,l)\neq(0,0)} K_{kl}\ketbra{j}{j'}K_{kl}^\dagger \nonumber \\
	&= \frac{1+(d^2-1)p}{d^2} L_{00} \ketbra{j}{j'} L_{00}^\dagger \nonumber \\
	& \quad + \frac{(1-p)}{d^2} \left( \sum_{k,l} L_{kl}\ketbra{j}{j'}L_{kl}^\dagger - L_{00} \ketbra{j}{j'} L_{00}^\dagger \right) \nonumber \\
	&= \frac{1 + (d^2-1)p - (1-p)}{d^2}\ketbra{j}{j'} \nonumber \\
	& \quad + \frac{(1-p)}{d^2} \sum_{k,l} \omega^{k(j-j')} \ketbra{j \oplus l}{j' \oplus l} \nonumber \\
	&= p \ketbra{j}{j'} + \frac{(1-p)}{d} \delta_{j,j'} \sum_l \ketbra{j \oplus l}{j' \oplus l} \nonumber \\
	&= p \ketbra{j}{j'} + \delta_{j,j'} (1-p) \frac{I}{d},
\end{align}
such that $\sum_{k,l} K_{kl} \rho K_{kl}^\dagger = p \rho + (1-p)I/d$ for any state $\rho$. This establishes that every element of $\mc{D}$ admits a set of SI Kraus operators in any basis.

Next, we take an arbitrary member $\mc{E}$ of $\bigcap_b \mc{SI}_b$ and show that it has the form of a depolarising channel. This proceeds in three stages. We show that the action on any pure state $\ket{\psi}$ is $\mc{E}(\ketbra{\psi}{\psi}) = p \ketbra{\psi}{\psi} + (1-p)I/d$, where $p$ can be a function of $\ket{\psi}$. We then narrow this down to $p$ being the same for all elements of a given basis, so that $\mc{E}(\ketbra{i}{i}) = p\ketbra{i}{i} + (1-p)I/d$ independent of $i$. Finally, we also deduce that $\mc{E}(\ketbra{i}{j}) = p \ketbra{i}{j}$ for any $i \neq j$. This suffices to determine the form of $\mc{E}$.

Take an arbitrary pure state $\ket{\psi}$, then there exists a basis $b = \{ \ket{i} \}_{i=0,\dots,d-1}$ with $\ket{0} = \ket{\psi}$. Now $\mc{E} \in \mc{SI}_b \Rightarrow \mc{E} \in \mc{MI}_b \Rightarrow \mc{E}(\ketbra{\psi}{\psi}) \in I_b$, thus
\begin{equation}
	\mc{E}(\ketbra{\psi}{\psi}) = \sum_i q_i \ketbra{i}{i}
\end{equation}
for some probabilities $q_i$. For $d=2$, we immediately have $\mc{E}(\ketbra{\psi}{\psi}) = p \ketbra{\psi}{\psi} + (1-p)I/2$ for some $p$ which is a function of $\ket{\psi}$. For $d>2$, take $i,j \neq 0$ such that $i \neq j$. Then there is a rotated basis $b'$ which is equal to $b$ except for the replacement of $\ket{i}, \ket{j}$ by $\ket{\pm_{ij}} := \frac{1}{\sqrt{2}}( \ket{i} \pm \ket{j})$, such that $\ketbra{\psi}{\psi} \in I_{b'}$. Now we have
\begin{align}
	\mc{E}(\ketbra{\psi}{\psi}) &= \left( \sum_{k \neq i,j} q_k \ketbra{k}{k} \right) \nonumber \\
	& + \frac{1}{2} \left[ (q_i+q_j)\ketbra{+_{ij}}{+_{ij}} + (q_i+q_j)\ketbra{-_{ij}}{-_{ij}} \right. \nonumber \\
	& \left. + (q_i-q_j)\ketbra{+_{ij}}{-_{ij}} + (q_i-q_j)\ketbra{-_{ij}}{+_{ij}} \right],
\end{align}
so $\mc{E}(\ketbra{\psi}{\psi}) \in I_{b'}$ only if $q_i = q_j$. Since $\mc{E} \in \mc{SI}_{b'}$, it follows that all $q_i$ for $i>0$ are equal to some $q$ independent of $i$.

Therefore we can write
\begin{align}
	\mc{E}(\ketbra{\psi}{\psi}) &= (1-q[d-1]) \ketbra{\psi}{\psi} + q \sum_{i>0} \ketbra{i}{i} \nonumber \\
	&= (1-q[d-1]) \ketbra{\psi}{\psi} + q \left( I - \ketbra{\psi}{\psi} \right) \nonumber \\
	&= (1-qd) \ketbra{\psi}{\psi} + (qd) I/d.
\end{align}
Evidently, for any $d$ we can write $\mc{E}(\ketbra{\psi}{\psi}) = p\ketbra{\psi}{\psi} + (1-p) I/d$. However, we must be careful that $p$ is in principle a function of $\ket{\psi}$. The following argument shows that it must in fact be a constant.

Choose an arbitrary basis $b = \{\ket{i}\}_{i=0,\dots,d-1}$. From above, for any $i$ we can write $\mc{E}(\ketbra{i}{i}) = p_i \ketbra{i}{i} + (1-p_i)I/d$. Using Lemma \ref{lem:unital_subset},
\begin{align}
	I &= \sum_i \mc{E}(\ketbra{i}{i}) \nonumber \\
	&= \sum_i p_i \ketbra{i}{i} + (1-p_i)I/d \nonumber \\
	&= \left( \sum_i p_i \ketbra{i}{i} \right) + \left(1 - \frac{\sum_i p_i}{d} \right) I,
\end{align}
which implies $p_i = \text{const.} =: p$. Note that $p$ still depends implicitly on the choice of basis.

For any pair $i \neq j$, construct the state $\ket{\psi} := \sqrt{a}\ket{i} + e^{i \phi} \sqrt{1-a}\ket{j}$ as a function of $a \in [0,1],\, \phi \in [0,2\pi]$. We know from above that
\begin{align}
	\mc{E}(\ketbra{k}{k}) &= p \ketbra{k}{k} + (1-p)I/d \quad \forall k, \\
	\mc{E}(\ketbra{\psi}{\psi}) &= r \ketbra{\psi}{\psi} + (1-r)I/d, \label{eqn:psi_depolarised}
\end{align}
where $r$ is implicitly a function of $a$ and $\phi$. From the definition of $\ket{\psi}$,
\begin{align}
	\mc{E}(\ketbra{\psi}{\psi}) &= a \left[ p\ketbra{i}{i} + (1-p)\frac{I}{d} \right] \nonumber \\
	& \quad + (1-a) \left[ p\ketbra{j}{j} + (1-p)\frac{I}{d} \right] \nonumber \\
	& \quad + \sqrt{a(1-a)} \mc{E}(e^{-i\phi}\ketbra{i}{j} + e^{i\phi}\ketbra{j}{i}),
\end{align}
and from (\ref{eqn:psi_depolarised}), we have
\begin{align}
	\mc{E}(\ketbra{\psi}{\psi}) &= r \left[ a\ketbra{i}{i} + (1-a)\ketbra{j}{j} \right. \nonumber \\
	& \quad + \left. \sqrt{a(1-a)} (e^{-i\phi}\ketbra{i}{j} + e^{i\phi}\ketbra{j}{i}) \right] \nonumber \\
	& \quad + (1-r)\frac{I}{d}.
\end{align}
We equate the two previous expressions and take the $\ketbra{i}{i}$ matrix element. Using the fact that $\mc{E} \in \mc{SI}_b$ implies $\mc{E}(\ketbra{i}{j})$ is fully off-diagonal, we obtain
\begin{align}
	ap + a(1-p)/d + (1-a)(1-p)/d &= ar + (1-r)/d \nonumber \\
	\Rightarrow p\left(a - \frac{1}{d}\right) &= r\left(a - \frac{1}{d}\right) \quad \forall a,\, \phi.
\end{align}
This immediately shows that $r = p \ \forall \phi,\, \forall a \neq 1/d$; by continuity of $\mc{E}$ we must have equality for all $a$.

Instead equating the off-diagonal parts of the two expressions for $\mc{E}(\ketbra{\psi}{\psi})$ (again using $\mc{E} \in \mc{SI}_b$), we have
\begin{equation}
	\sqrt{a(1-a)} \mc{E}(e^{-i\phi}\ketbra{i}{j} + \text{h.c.}) = r \sqrt{a(1-a)}(e^{-i\phi}\ketbra{i}{j} + \text{h.c.})
\end{equation}
for all $\phi$, where h.c.\ stands for the hermitian conjugate of $e^{-i\phi}\ketbra{i}{j}$. Choosing any $a \neq 0,1$ and using $r=p$,
\begin{align}
	\phi=0:  \quad & \mc{E}(\ketbra{i}{j}) + \mc{E}(\ketbra{j}{i}) = p(\ketbra{i}{j} + \ketbra{j}{i}) \\
	\phi=\pi/2:  \quad & -i\mc{E}(\ketbra{i}{j}) + i\mc{E}(\ketbra{j}{i}) = p(-i\ketbra{i}{j} + i\ketbra{j}{i}),
\end{align}
so that $\mc{E}(\ketbra{i}{j}) = p \ketbra{i}{j}$. Together with $\mc{E}(\ketbra{i}{i}) = p\ketbra{i}{i} + (1-p)I/d$, this is all we need to conclude that $\mc{E}(\rho) = p \rho + (1-p)I/d$ for any state $\rho$.

\bibliography{using_coherence}
\end{document}